\documentclass[aps,prb,preprint,superscriptaddress,nobibnotes]{revtex4-1}
\usepackage{amsmath}
\usepackage{bbm}
\usepackage{commath}
\usepackage{graphicx}
\usepackage{color}
\usepackage{indentfirst}
\usepackage{setspace}
\usepackage{tabularx}
\usepackage{multirow}
\usepackage{subfigure}
\usepackage{float}

\begin{document}

\title{Re-entrant melting of sodium, magnesium and aluminum \\and the general trend} 

\author{Qi-Jun Hong}
\email[e-mail:]{ qhong@alumni.caltech.edu}
\affiliation{School of Engineering, Brown University, Providence, Rhode Island 02912, USA}

\author{Axel van de Walle}
\affiliation{School of Engineering, Brown University, Providence, Rhode Island 02912, USA}
\date{\today}

\begin{abstract}
Re-entrant melting (in which a substance's melting point starts to decrease beyond a certain pressure) is believed to be an unusual phenomenon. 
Among the elements, it has so far only been observed in a very limited number of species, \textit{e.g.}, the alkali metals.
Our density functional theory calculations reveal that this behavior actually extends beyond alkali metals to include magnesium, which also undergoes re-entrant melting, though at the much higher pressure of $\sim$300 GPa.
We find that the origin of re-entrant melting is the faster softening of interatomic interactions in the liquid phase than in the solid, as pressure rises. 
We propose a simple approach to estimate pressure-volume relations and show that this characteristic softening pattern is widely observed in metallic elements. 
We verify this prediction in the case of aluminum by finding re-entrant melting at $\sim$4000 GPa.
These results suggest that re-entrant melting may be a more universal feature than previously thought.
\end{abstract}
\maketitle

Re-entrant melting is generally considered an unusual phenomenon \cite{Schwegler07,Giaquinta05} that is associated with a negative slope of the melting temperature versus pressure line, or the melting curve.
An ``ordinary" melting curve, rising from low temperatures and pressures, may eventually invert its trend at some maximum temperature, with a change of slope from positive to negative values for increasing pressures. 
The resulting topology of the melting curve gives rise to re-entrant melting: upon compressing the liquid at temperatures lower than the maximum melting temperature, one observes a liquid-solid-liquid sequence of phases. In other words, the liquid phase, which is stable at low pressures, re-enters at higher pressures.

Most melting curves exhibit positive melt slopes, while the occurrence of negative slopes and re-entrant melting is overall far less frequent and varies widely depending on the type of materials.
On the one hand, negative melt slopes are not uncommon among nonmetals and complex chemical systems, \textit{e.g.}, silicon, gallium, antimony, and water \cite{Young91}. 
According to the Clausius-Clapeyron relation, this negative slope is associated with open crystalline structure and large volume in the solid phase, which leads to higher density in the liquid phase and negative volume change upon melting (which we will discuss in detail later).
On the other hand, metals are more close-packed, and hence re-entrant melting is observed so far in only a very limited number of metals, \textit{e.g.}, the alkali metals \cite{Parrinello12}, while most metals exhibit melting curves that increase indefinitely as pressure rises \cite{Young91,Alfe99,Erran01,Dewaele10,Burakovsky10,Bouchet09,Hieu13,Stutzmann15,Hrubiak17,Boccato17,Burakovsky16,Cazorla16,Errandonea18,Anzellini19,Errandonea19}.
Intuitively, atomic interactions become overwhelmingly repulsive at high pressure, so it is fair to expect any deviation from a crystalline order that reduces packing efficiency to become less favorable in terms of enthalpy, which results in high melting point at high pressure.
Historically, the wide acceptance of the Lindemann and Gr\"uneisen laws \cite{Lind10,Grun26,Gilv56} and the Simon-Glatzel equation \cite{Simo29} leads to a general perception that ``normal" melting curves rise indefinitely with increasing pressure.
A common melting curve fitting based on the Lindemann law typically involves extrapolation to high pressure region from low pressure data, presuming same melting properties regardless of change in pressure,
\begin{equation}
\frac{\dif \ln{T_m}}{\dif P} = \frac{2(\gamma_m-1/3)}{B_m},
\end{equation}
where $T_m$ is melting temperature, $P$ is pressure, $\gamma_m$ is the Gr\"uneisen melting parameter and $B_m$ is the bulk modulus of solid.
When $\gamma_m>1/3$, which is generally true for most metals at low pressures \cite{Bura04}, the right-hand side of the equation is positive and thus melting temperature increases indefinitely with pressure.
In the well-known Simon-Glatzel empirical equation of melting, temperature monotonically increases as seen in its analytic form,
\begin{equation}
T_m = T_0 ( 1 + (P-P_0) / a)^b,
\end{equation}
where $a$ and $b$ are fitting parameters and $(T_0, P_0)$ are reference melting temperature and pressure.
To model decreasing melting curves requires more sophisticated models, such as the Kechin equation \cite{Kech01}.

In this work, we employ density functional theory \cite{Kohn64,Kohn65,Jones89} to calculate melting curves of metals, from which we detect and locate re-entrant melting.
We discover that re-entrant melting occurs far more widely than is generally recognized.
To calculate melting temperatures of metals under various pressure conditions, we use an efficient extension of the coexistence method \cite{Hong13} and its implementation in the SLUSCHI code \cite{Hong16}, based on density functional theory molecular dynamics.
This highly efficient method makes it possible to perform, directly from first principles, dozens of expensive melting point calculations on sodium, magnesium, and aluminum, which are otherwise considered prohibitively expensive, given the various combination of systems and pressures. The method runs solid-liquid coexisting simulations on small-size systems, and the melting temperatures are rigorously inferred based on statistical analysis of the fluctuations and probability distributions in the systems \cite{Hong13,Hong16}.
The accuracy (typically with an error smaller than 100K), robustness and efficiency of the method have been demonstrated in a range of materials \cite{Hong13,Hong16,Hong15,Hong_HfTaC,Mil15,Guren17,Addington16}.

Density functional theory calculations were performed by the Vienna Ab-initio Simulation Package (VASP) \cite{VASP}, with the projector-augmented-wave (PAW) \cite{BLOCHL94} implementation and the generalized gradient approximation (GGA) for exchange-correlation energy, in the form known as Perdew,
Burke, and Ernzerhof (PBE) \cite{PBE96}.
Since the simulations were performed at high pressure conditions, we used accurate pseudopotentials where the semi-core $s$ and $p$ states were treated as valence states.
The accuracy of the PAW pseudopotentails, even under extreme pressure conditions where re-entrant melting occurs, is further validated by comparison with the WIEN2K code \cite{WIEN2K} based on the full-potential linearized augmented plane-wave (LAPW) method (see Supplemental Material \cite{SM,VASP96-2,VASP99,Nose84,Nose84-2,Hoover85,Klein92}).

\begin{figure}
\centering
\includegraphics[width=0.48\textwidth]{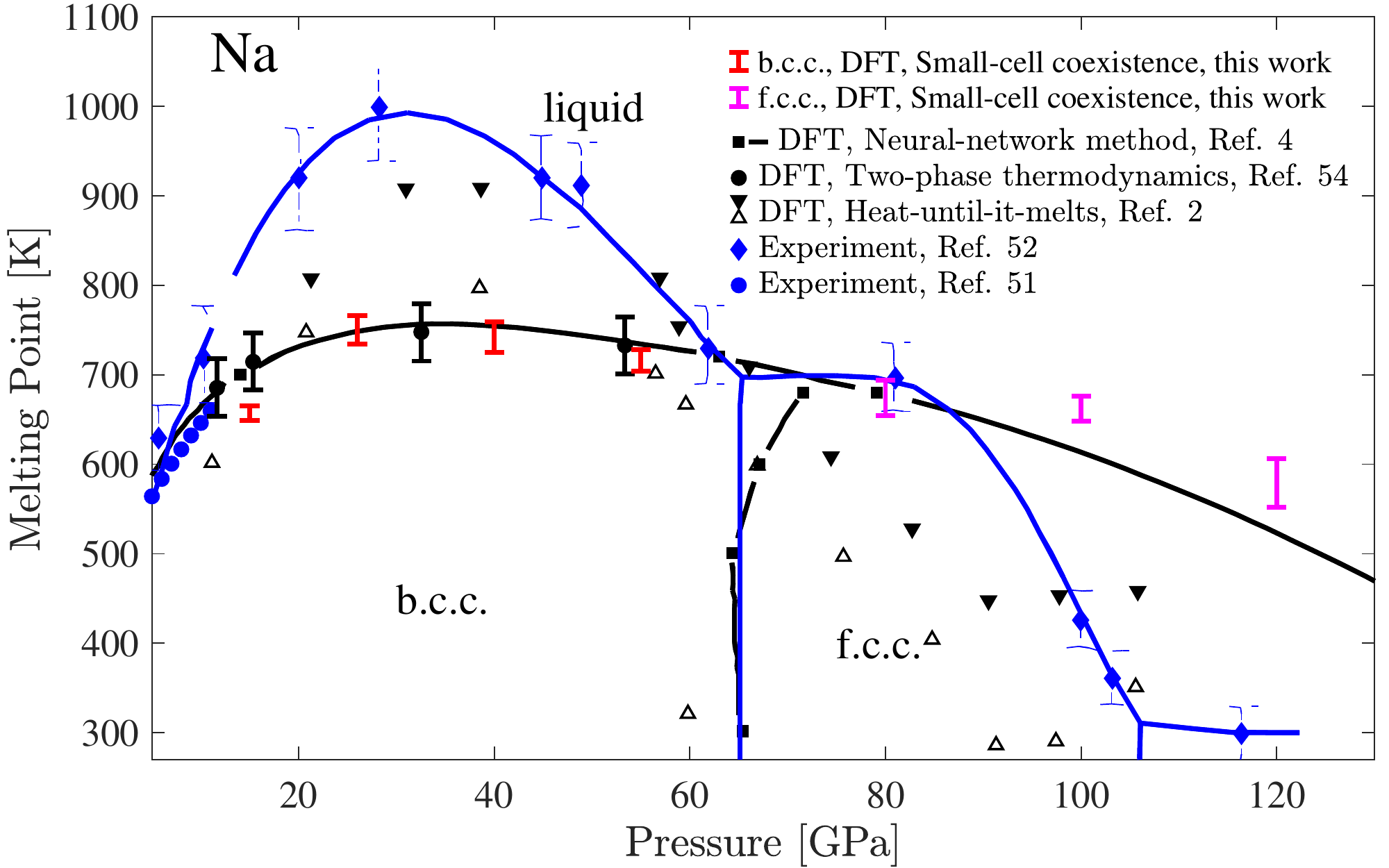}
\caption{\label{Na}
Melting curve of sodium. Our calculated melting curve (in red and magenta) agree closely with two other DFT calculations (Ref. \onlinecite{Parrinello12} and \onlinecite{Desjarlais13}).}
\end{figure}

Our investigation starts with sodium, an alkali element well known for re-entrant melting.
Despite being a prototype of simple metal at ambient conditions, sodium exhibits unexpected complexity at high pressure \cite{Naumov15, Rousseau11, Naumov15-2}. 
Early experimental measurements of its melting curve \cite{Luedemann68,Mirwald76,Boehler85}, while many closely agree with each other, obtained only limited pressure up to 12 GPa and thus fell short of finding re-entrant melting.
Later, experiments by Gregoryanz \textit{et al.} \cite{Mao05} extended these investigations to much higher pressure and discovered that the sodium melting curve reaches a maximum at around 30 GPa followed by a pressure-induced drop, which extends to nearly room temperature at $\sim$120 GPa and over the stability regions of three solid phases. 
Though Gregoryanz's work is well accepted by the community, comparison with its predecessors reveals considerable discrepancy:
its melting temperatures are noticeably higher and, with a significantly steeper slope, its melting curve starts to diverge from others as pressure increases, as shown in Fig. \ref{Na}.

There have been several pieces of computational work supporting the occurrence of re-entrant melting, as summarized in Fig. \ref{Na}.
Using an \textit{ab initio} quality neural-network potential, Eshet \textit{et al.} \cite{Parrinello12} calculated a melting curve through the free energy method \cite{Sugino95, Alfe99}. This melting curve was later confirmed by Desjarlais \cite{Desjarlais13} based on DFT calculations and the two-phase thermodynamics method \cite{Lin03}.
Both DFT-level investigations, regardless of different approaches employed to compute melting temperature, agreed closely with melting curves of early experimental results \cite{Luedemann68,Mirwald76,Boehler85} in low pressure region.
At high pressures, the two computational studies nevertheless gave melting temperatures that were widely different from Gregoryanz's: melting temperature maxima were lower than Gregoryanz's finding by as much as 250 K.
Raty \textit{et al.} \cite{Schwegler07} employed the ``heat-until-it-melts" approach and obtained a melting curve that lies between the two sides. 

Employing the small-size coexistence method \cite{Hong13} and the SLUSCHI code \cite{Hong16} that we recently developed,  we provide independent corroboration from another perspective and help to resolve the remaining dispute.
As shown in Fig. \ref{Na}, we confirm the existence of re-entrant melting near 750 K and 35 GPa.
Our results agree closely with Eshet's \cite{Parrinello12} and Desjarlais' \cite{Desjarlais13} DFT calculations, as well as early experimental measurements \cite{Luedemann68,Mirwald76,Boehler85} in low pressure region.
We note that these three DFT-based results, irrespective of different computational approaches employed to compute melting points, are remarkably consistent.
Hence it is likely that they together well establish the DFT melting curve of sodium.
While the ``heat-until-it-melts" results are slightly higher, the method is known to inherently overestimate melting point \cite{Alfe11} and we thus view the results as upper boundaries.
It is not clear why Gregoryanz's experimental melting curve \cite{Mao05}, though widely accepted, singled itself out as being significantly higher than the DFT melting curve. 
Both computation and experiment have flaws that may cause the discrepancy, \textit{e.g.}, inaccurate DFT exchange-correlation functional, and challenging experimental conditions, such as sodium being highly reactive. 
We note that early experimental measurements \cite{Luedemann68,Mirwald76,Boehler85} seem to favor the DFT melting curve, but the limited pressure in these studies prevent them from resolving the dispute conclusively.
We believe that this issue deserves more experimental investigations.

\begin{figure}
\centering
\includegraphics[width=0.48\textwidth]{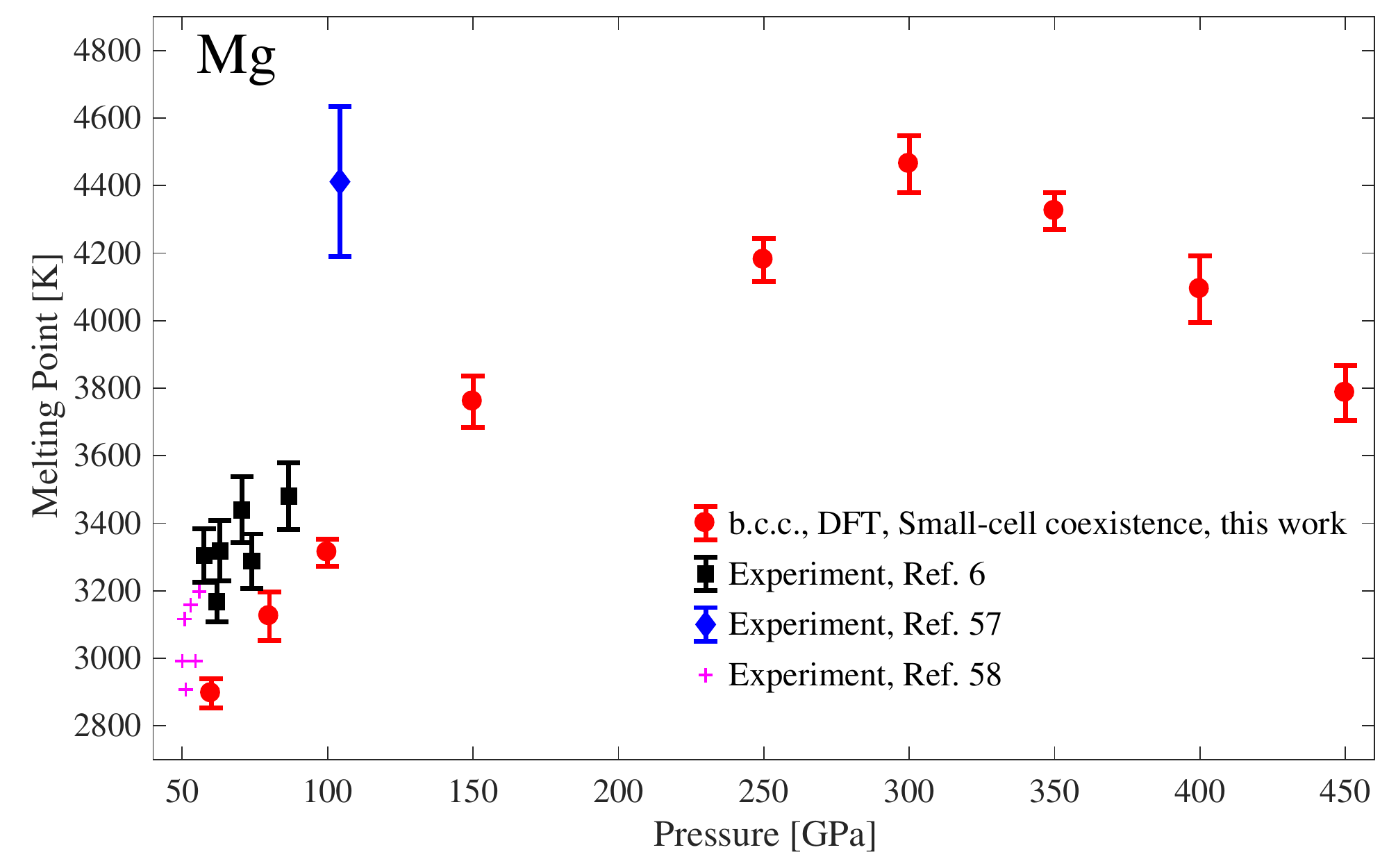}
\caption{\label{Mg}
Melting of magnesium. Melting curve maximum is discovered in magnesium at $\sim$300 GPa and 4500 K.}
\end{figure}

Despite being an intriguing phenomenon, re-entrant melting is not well studied among elements beyond alkali metals, due to challenging experimental condition at high pressures. 
We here extend our theoretical investigation past sodium, and discover, for the first time, that re-entrant melting also exists in magnesium, the next period three element.
However, it occurs at a much higher pressure.
As illustrated in Fig. \ref{Mg}, our calculations reveal that re-entrant melting of magnesium takes place at $\sim$300 GPa and 4500 K, which is  about one order of magnitude higher than that of sodium. 
Magnesium undergoes a phase transition from h.c.p. to b.c.c. at 50 GPa \cite{Erran01,Stinton14}, and the b.c.c. phase is stable throughout the high-pressure region, according to our simulations.
While no experimental data is available at the extreme condition near the re-entrant point, at relatively low pressures below 100 GPa our results are mostly consistent with two pieces of experimental work by Errandonea \textit{et al.} \cite{Erran01} and by Urtiew and Grover \cite{Urtiew77}, with our computational melting temperatures near the lower boundary of the experimental data.
We note that another experimental work by Stinton \textit{et al.} \cite{Stinton14} gives substantially higher melting point at 100 GPa. 
In Errandonea's work \cite{Erran01}, melting is determined from the properties on the surface. In Stinton's work \cite{Stinton14}, crystal structure is determined by X-ray that goes through the sample, while temperature is still measured from the surface, as there is no other way to measure it. 
There can be temperature gradients, being the surface hotter than the inner part of the sample, then it is reasonable to get higher melting temperature when measuring melting using X-ray diffraction. This can lead to an overestimation of the melting temperature.

\begin{figure}
\centering
\includegraphics[width=0.23\textwidth]{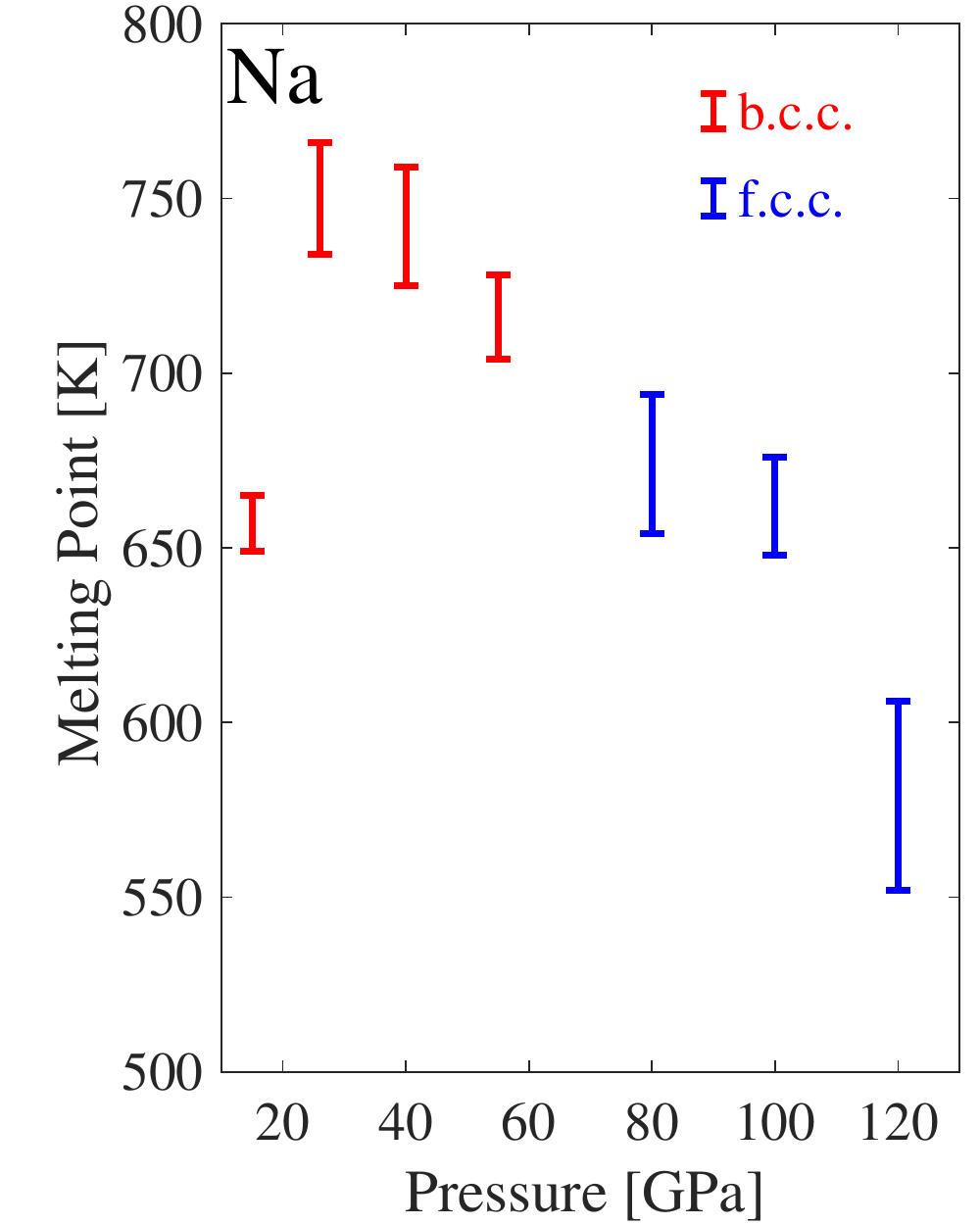}
\includegraphics[width=0.23\textwidth]{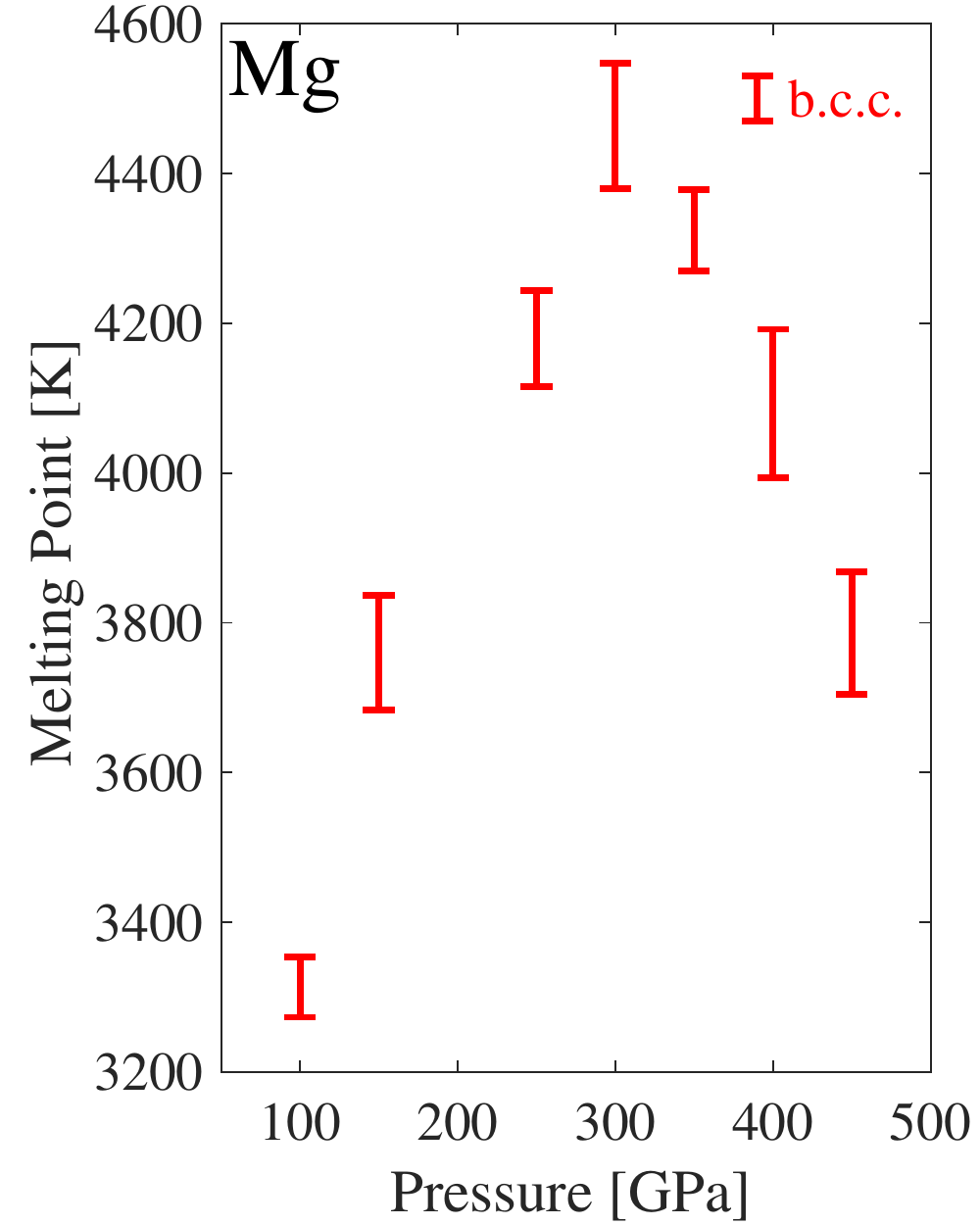}\\
\includegraphics[width=0.23\textwidth]{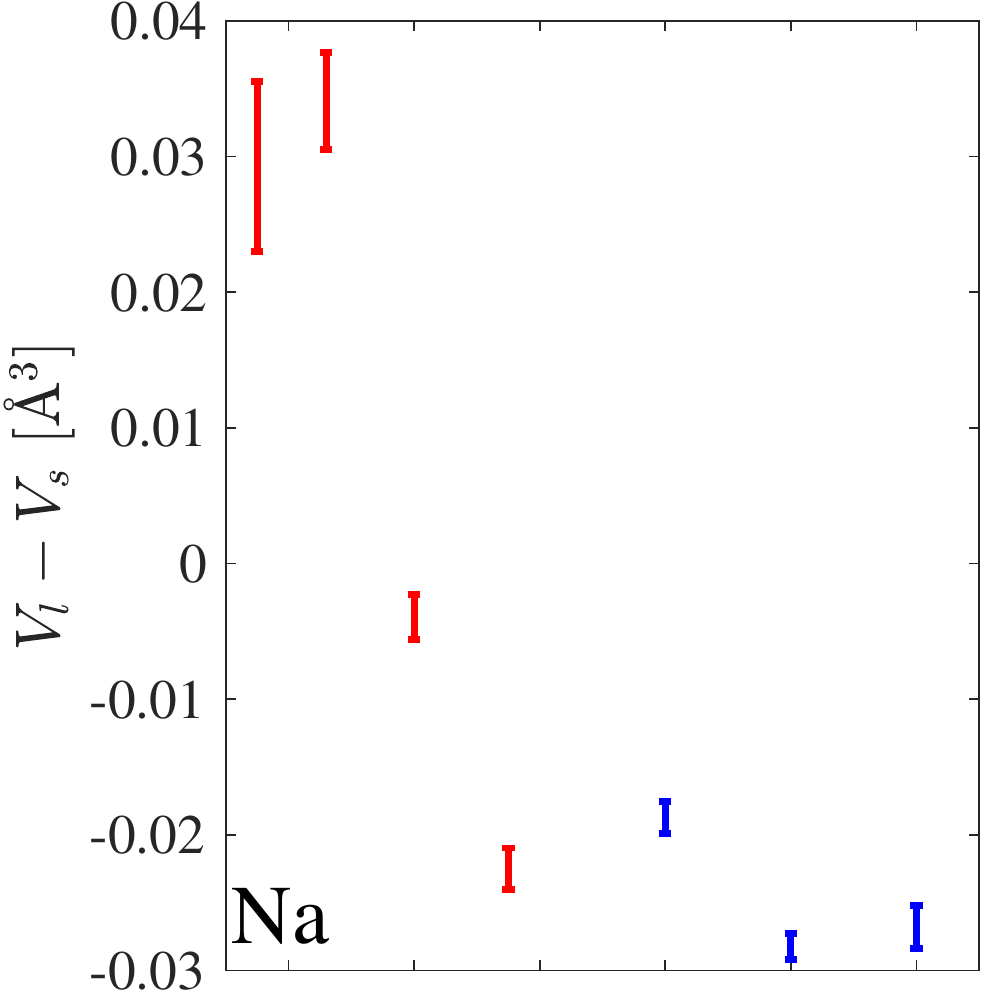}
\includegraphics[width=0.23\textwidth]{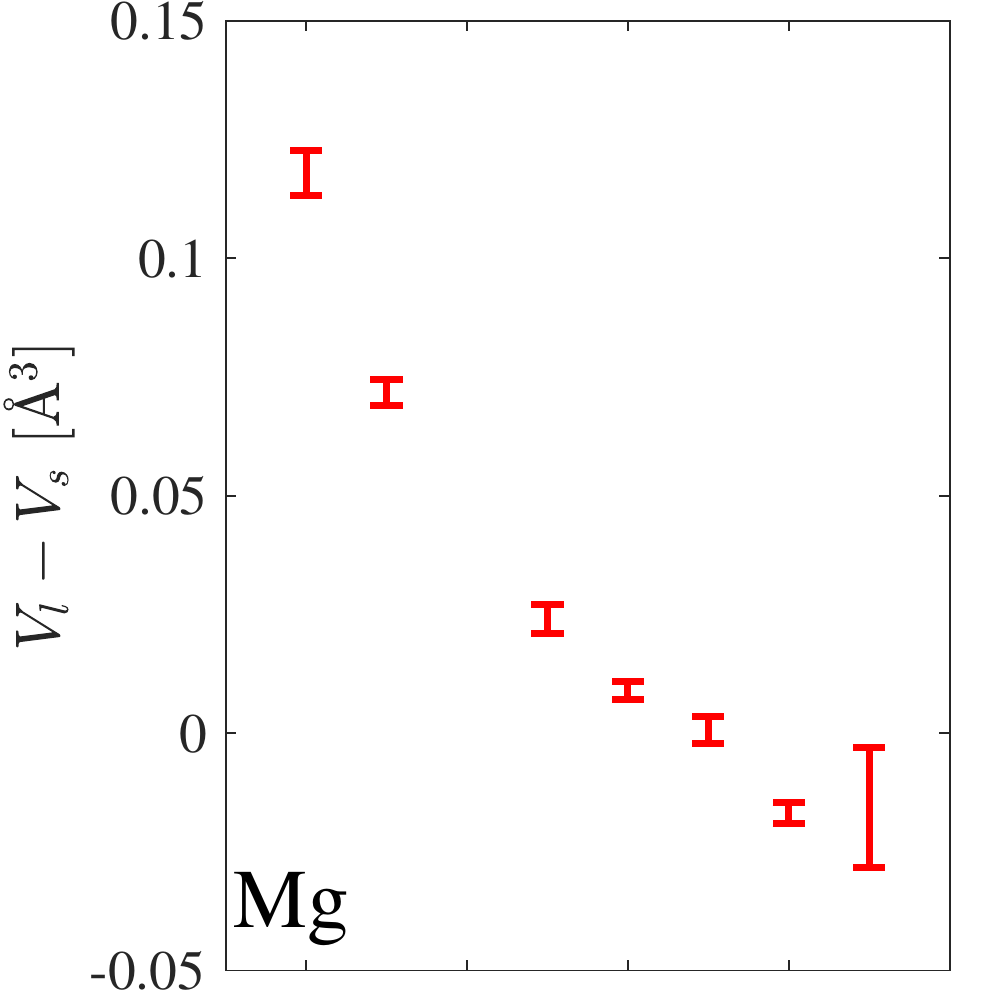}
\caption{\label{NaMg}
Melting temperatures and volume changes upon melting of Na and Mg under high pressures. The change of sign in $\Delta V$  coincides with  the occurrence of the melting curve maximum.}
\end{figure}

By probing into similarities and differences between sodium and magnesium, we aim to understand the physics underlying re-entrant melting.
As it has been widely acknowledged\cite{Giaquinta05,Schwegler07,Parrinello12,Mao05}, a negative melt slope is linked to a decrease in specific volume upon melting.
According to the Clausius-Clapeyron relation, the slope of a melting curve is determined by 
\begin{equation}
\frac{\dif P}{\dif T_m} = \frac{\Delta H}{T_m\Delta V},
\end{equation}
where $P$ is pressure, $T_m$ is melting temperature, $\Delta V=V_l-V_s$ is difference between solid and liquid in specific volumes and $\Delta H$ is specific heat of fusion.
A negative melting line slope ${\dif T_m}/{\dif P}$ occurs when volume change $\Delta V$ is negative, \textit{i.e.}, the liquid phase has a smaller specific volume than the solid does.
Indeed, the occurrence of negative melt slope in several nonmetals can be attributed to their open crystalline structure and large volume in solid phase, which leads to negative volume change upon melting.
For sodium and magnesium, our predicted change of sign in $\Delta V$ exactly coincides with  the occurrence of the melting curve maximum, as shown in Fig. \ref{NaMg}, thus providing additional corroboration of the re-entrant behavior.

There is an ongoing debate \cite{Giaquinta05,Schwegler07,Parrinello12} over the origin of this negative volume change.
Volume can be decomposed into two factors: (1) coordination number and (2) pairwise distance between nearest neighbors. Both higher coordination number (or packing efficiency) and shorter radial distance can lead to smaller volume.
On the one hand, the formation of low-symmetry structures with low-coordination numbers that frequently observed in dense lithium and sodium complex structures gives rise to large volume of the solid phase \cite{Naumov15, Rousseau11, Naumov15-2}.
For Mg, the solid undergoes a h.c.p.-b.c.c. phase transition at $\sim$50 GPa \cite{Erran01,Stinton14} and reduces the coordination number from 12 to 8, which is closer to the more open coordination of the melt.
On the other hand, Raty \textit{et al.} \cite{Schwegler07} found that the small volume of liquid is induced by ``uneven" softening of effective intermolecular interactions, which occurs at a faster rate in the liquid than in the solid for growing pressures.
As a result of the smaller radial distance, the liquid phase shrinks faster than the solid. 

\begin{figure}
\centering
\includegraphics[width=0.48\textwidth]{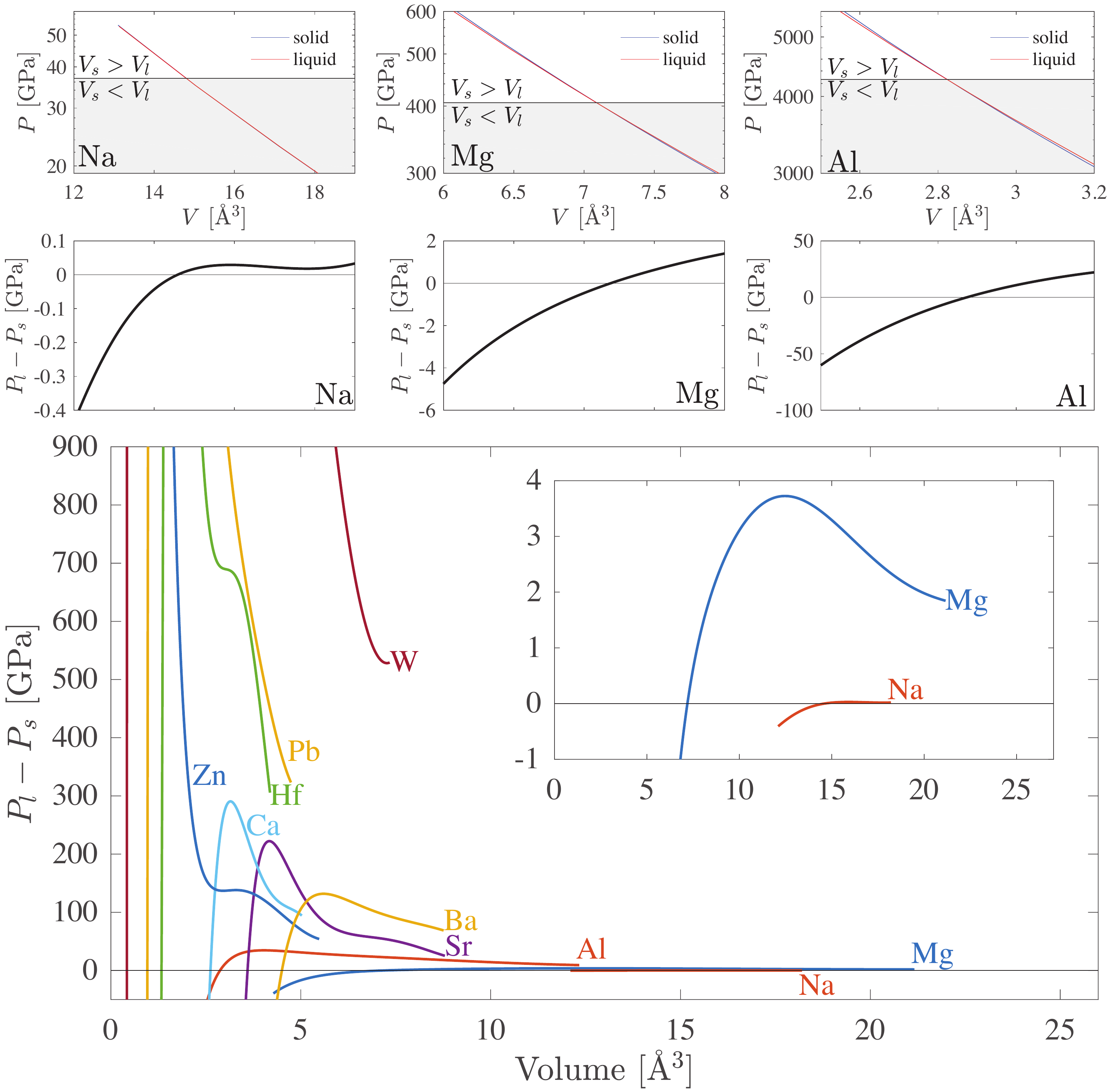}
\caption{\label{DelP}
Pressure difference between solid and liquid phases, versus volume. 
Results are based on analysis of one randomly sampled snapshot from each phase of each metal.
At a large volume, $P_l$ is always higher than $P_s$. In other words, $V_l > V_s$ if the pressure is low.
This relation inverts at small volumes and high pressures, as $P_l-P_s$ turns negative, which leads to negative volume change and re-entrant melting.
}
\end{figure}

Our investigation into this issue supports the second reasoning.
Our analysis of b.c.c. and liquid structures of magnesium (see Supplemental Material \cite{SM,VASP96-2,VASP99,Nose84,Nose84-2,Hoover85,Klein92}) does not detect significant difference in coordination number, while, following the first reasoning, the relatively low coordination number in b.c.c. should result in a significant increase upon melting. This contradiction suggests that coordination number does not play a significant role in the volume decrease. 
Furthermore, in the sodium phase diagram, melting temperature continues to decrease with pressure, even in the region where the f.c.c. phase is stable (Fig. \ref{NaMg}). Since the close-packed f.c.c. structure maximizes the coordination number, it is not sensible to attribute the volume decrease to coordination number.
To further distinguish these two factors, we note that we can adopt a simple procedure to change radial distances while leaving coordination number fixed.
We first randomly sample one snapshot from each MD trajectory of the solid and the liquid, and we use this snapshot to represent each phase.
We then uniformly scale the lattice vectors and compress the structure to study and estimate the pressure-volume relation, while changing only the radial distances. This process does not change the coordination numbers, since we keep the atomic fractional coordinates untouched.
For both sodium and magnesium, our calculations find that volume difference between the solid and the liquid snapshots turns negative at sufficiently high pressures, as plotted in Fig. \ref{DelP}, which is strong evidence in favor of shorter radial distance and faster potential softening in the liquid phase.
Moreover, the location where the two curves cross is fairly consistent with the actual melting point maximum, when compared to Fig. 1. 
Therefore, in addition to revealing the nature of re-entrant melting, this simple and approximated procedure of scaling is valuable to quickly screen a material for re-entrant melting and to locate its position.
We note that melting point calculations are much more expensive than static calculations of pressure and energy. While it can be prohibitively expensive to look for re-entrant melting from scratch without clues regarding its pressure condition, searching for negative volume change is feasible and it enables us to rapidly locate re-entrant melting.

\begin{figure}
\centering
\includegraphics[width=0.45\textwidth]{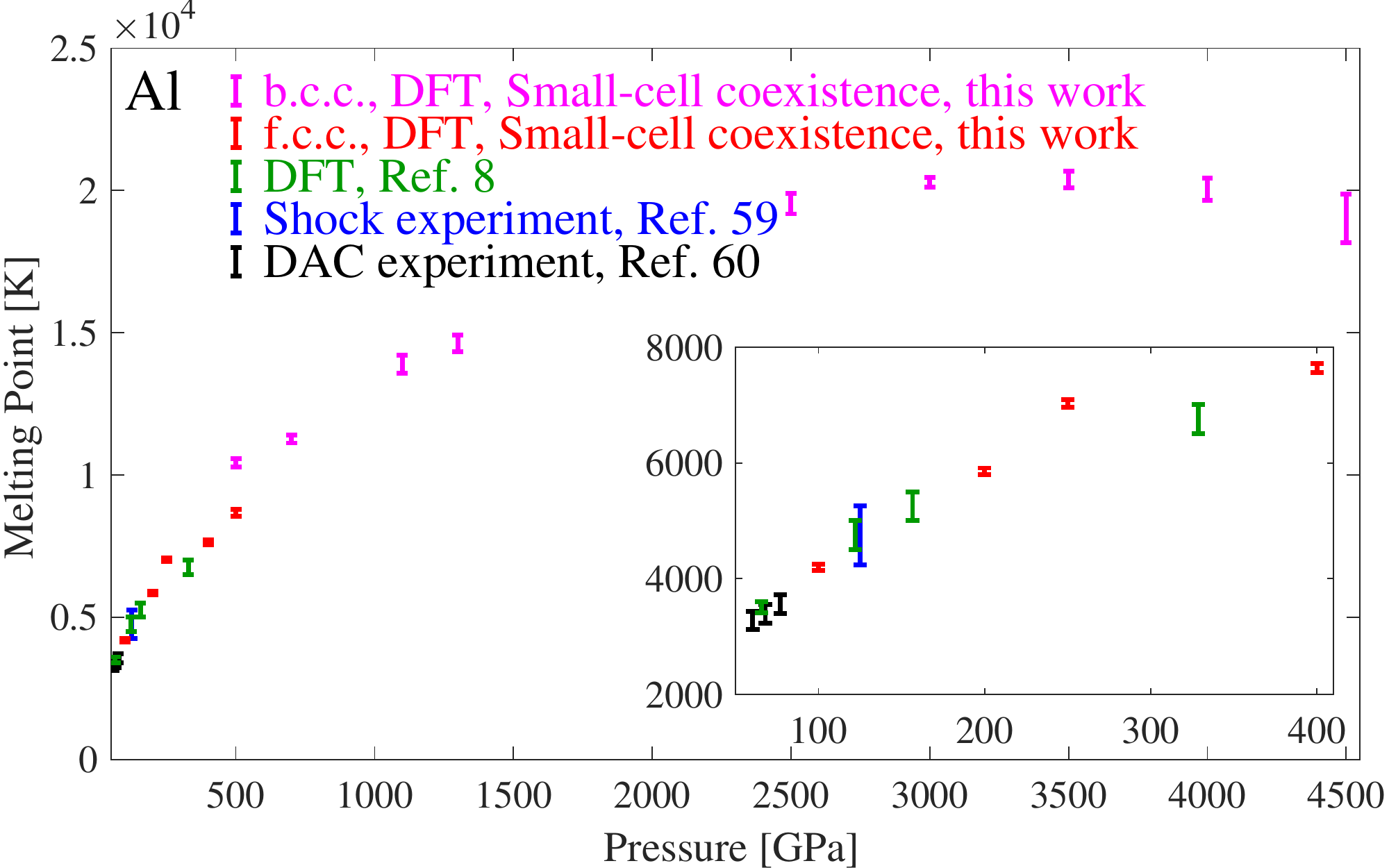}
\caption{\label{Al}
Melting curve and re-entrant melting of Al under high pressure. We discover melting point maximum of aluminum at $\sim$3500 GPa and 20000 K.}
\end{figure}

The existence of re-entrant melting in both sodium and magnesium raises the possibility that re-entrant melting is a universal feature of all metals.
Here we investigate the general trend of re-entrant melting in various elements across the periodic table.
Employing the aforementioned scaling method, we carry out quick tests on a wide variety of metals to see whether similar liquid-state potential softening exists.
As shown in Fig. \ref{DelP}, our calculations confirm indeed stronger softening of interatomic potential in the liquid phase than in the solid for many metals, including calcium, strontium, aluminum, titanium, lead, hafnium, etc. 
The faster potential softening in liquids indicates that negative volume change, and thus re-entrant melting as well, may also occur in these metals at high pressures.
For several metals, \textit{e.g.}, zinc, negative volume change is not achieved due to the limit we can compress the structures at extremely high pressure and the failure of pseudopotentials. However, we do not totally rule out the possibility.
While it is still premature to state that re-entrant melting is a universal behavior of all metals, its surprisingly wide occurrence points to a real possibility and extends our understanding on this topic.

Although it is very expensive to perform first-principles melting curve calculations on all the metals found, we select one metal, aluminum, to ascertain our findings. With a large bulk modulus and high melting temperature, aluminum would initially appear to be a poor candidate for re-entrant melting. However, from the estimated pressure condition of re-entrant melting in Fig. \ref{DelP}, we perform melting point calculations in the pressure range and we confirm that re-entrant melting indeed exists in aluminum, though at an extremely high pressure and temperature. As shown in Fig. \ref{Al}, b.c.c. aluminum achieves a melting temperature maximum at $\sim$3500 GPa and 20000 K. 
At relatively low pressure and temperature, our computational melting points agree closely with previously reported experiments \cite{Boehler97,Shaner84}, as well as DFT data based on solid-liquid coexistence simulations \cite{Bouchet09}.
In addition to the discovery of re-entrant melting, we also find that b.c.c. aluminum is more stable than f.c.c. starting at around 500 GPa, and the b.c.c. phase persists at high pressures, partially corroborating recent b.c.c.-f.c.c. phase transition findings from both computation \cite{Sjostrom16} and experiment \cite{Polsin17}.
While pressures up to 5000 GPa may be difficult to reach in laboratory settings, such pressures do occur in the Universe.
For example, the inner core of Jupiter is roughly 3000-4500 GPa \cite{Elkins06}, which is comparable to the pressure range.

We should point out that the small-size coexistence method has the built-in ability to identify the correct structure of the solid  because the structure with the lowest free energy can easily nucleate at the solid-liquid interface even if the wrong solid structure was initially assumed. We have observed this to happen in the Al case: even if the simulation is initially setup with an f.c.c. solid, a b.c.c. solid spontaneously forms above a certain pressure ($\sim$500 GPa).

To summarize, we investigate the phenomenon of re-entrant melting based on density functional theory calculations. 
We discover, for the first time, that magnesium and aluminum also have this feature in their phase diagrams, similar to sodium.
We confirm that the origin of re-entrant melting is the faster softening of interatomic potentials and hence smaller volume in the liquid phase than in the solid, as pressure rises.
We propose a quick approach to estimate the pressure-volume relation and show that this phenomenon is widely observed in metals, and hence raises the possibility that re-entrant melting is a universal property of materials.

\section*{Acknowledgement}
The authors thank Dr. Errandonea for discussion.
This research was supported by National Science Foundation under grant DMR-1835939, by Office of Naval Research under grants N00014-14-1-0055 and N00014-17-1-2202, and by Brown University through the use of the facilities at its Center for Computation and Visualization. 
This work uses the Extreme Science and Engineering Discovery Environment (XSEDE), which is supported by National Science Foundation grant number ACI-1548562.


\begin{thebibliography}{}

\bibitem{Giaquinta05}
P. V. Giaquinta and F. Saija, Re-entrant melting in the Gaussian-core model: the entropy imprint, Chem. Phys. Chem. {\bf 6}, 1768-1771 (2005).

\bibitem{Schwegler07}
J.-Y. Raty, E. Schwegler, and S. A. Bonev, Electronic and structural transitions in dense liquid sodium, Nature {\bf 449}, 448-451 (2007).

\bibitem{Young91}
D. A. Young, \textit{Phase Diagrams of the Elements}, (University of California Press, Berkeley, 1991).

\bibitem{Parrinello12}
H. Eshet, R. Z. Khaliullin, T. D. Kuhne, J. Behler, and M. Parrinello, Microscopic origins of the anomalous melting behavior of sodium under high pressure, Phys. Rev. Lett. {\bf 108}, 115701 (2012).

\bibitem{Alfe99}
D. Alf\`{e}, M. Gillan, and G. Price, The melting curve of iron at the pressures of the Earth's core from \textit{ab initio} calculations, Nature \textbf{401}, 462-464 (1999).
\bibitem{Erran01}
D. Errandonea, R. Boehler, and M. Ross, Melting of the alkaline-earth metals to 80 GPa, Phys. Rev. B {\bf 65}, 012108 (2001).
\bibitem{Dewaele10}
A. Dewaele, M. Mezouar, N. Guignot, and P. Loubeyre, High melting points of tantalum in a laser-heated diamond anvil cell, Phys. Rev. Lett. {\bf 104}, 255701 (2010).
\bibitem{Burakovsky10}
L. Burakovsky, S. P. Chen, D. L. Preston, A. B. Belonoshko, A. Rosengren, A. S. Mikhaylushkin, S. I. Simak, and J. A. Moriarty, High-pressure--high-temperature polymorphism in Ta: resolving an ongoing experimental controversy, Phys. Rev. Lett. {\bf 104}, 255702 (2010).
\bibitem{Bouchet09}
J. Bouchet, F. Bottin, G. Jomard, and G. Z\'{e}rah, Melting curve of aluminum up to 300 GPa obtained through \textit{ab initio} molecular dynamics simulations, Phys. Rev. B {\bf 80}, 094102 (2009).
\bibitem{Hieu13}
H. K. Hieu and N. N. Ha, High pressure melting curves of silver, gold and copper, AIP Advances {\bf 3}, 112125 (2013).
\bibitem{Stutzmann15}
V. Stutzmann, A. Dewaele, J. Bouchet, F. Bottin, and M. Mezouar, High-pressure melting curve of titanium, Phys. Rev. B {\bf 92}, 224110 (2015).
\bibitem{Hrubiak17}
R. Hrubiak, Y. Meng, and G. Shen, Microstructures define melting of molybdenum at high pressures, Nat. Commun. \textbf{8}, 14562 (2017).

\bibitem{Boccato17}
S. Boccato, R. Torchio, I. Kantor, G. Morard, S. Anzellini, R. Giampaoli, R. Briggs, A. Smareglia, T. Irifune, and S. Pascarelli, The melting curve of nickel up to 100 GPa explored by XAS, J. Geophys. Res. Solid Earth \textbf{12}, 9921 (2017).
\bibitem{Burakovsky16}
L. Burakovsky, N. Burakovsky, M. J. Cawkwell, D. L. Preston, D. Errandonea, and S. I. Simak, Ab initio phase diagram of iridium, Phys. Rev. B \textbf{94}, 094112 (2016).
\bibitem{Cazorla16}
C. Cazorla, S. G. MacLeod, D. Errandonea, K. A. Munro, M. I. McMahon, and C. Popescu, Thallium under extreme compression, J. Phys. Condens. Matter \textbf{28}, 445401 (2016).
\bibitem{Errandonea18}
D. Errandonea, S. G. MacLeod, J. Ruiz-Fuertes, L. Burakovsky, M. I. McMahon, C. W. Wilson, J. Iba{\~{n}}ez, D. Daisenberger, and C. Popescu, High-pressure/high-temperature phase diagram of zinc, J. Phys. Condens. Matter \textbf{30}, 295402 (2018).
\bibitem{Anzellini19}
S. Anzellini, V. Monteseguro, E. Bandiello, A. Dewaele, L. Burakovsky, and D. Errandonea, In situ characterization of the high pressure -- high temperature melting curve of platinum, Sci. Rep. \textbf{9}, 13034 (2019).
\bibitem{Errandonea19}
D. Errandonea, S. G. MacLeod, L. Burakovsky, D. Santamaria-Perez, J. E. Proctor, H. Cynn, and M. Mezouar, Melting curve and phase diagram of vanadium under high-pressure and high-temperature conditions, Phys. Rev. B \textbf{100}, 094111 (2019).

\bibitem{Lind10}
F. A. Lindemann, The calculation of molecular vibration frequencies, Phys. Z. {\bf 11}, 609 (1910).
\bibitem{Grun26}
E. Gr\"uneisen, in \textit{Handbuch der Physik} (Verlag Julius Springer, Berlin, 1926), pp. 1-59.
\bibitem{Gilv56}
J. J. Gilvarry, The Lindemann and Gr\"uneisen laws, Phys. Rev. {\bf 102}, 308-316 (1956).

\bibitem{Simo29}
F. E. Simon and G. Z. Glatzel, Z. Anorg. Allg. Chem. {\bf 178}, 309 (1929).

\bibitem{Bura04}
L. Burakovsky and D. L. Preston, Analytic model of the Gr\"{u}neisen parameter all densities, J. Phys. Chem. Solids {\bf 65}, 1581-1587 (2004).

\bibitem{Kech01}
V. V. Kechin, Melting curve equations at high pressure, Phys. Rev. B {\bf 65}, 052102 (2001).

\bibitem{Kohn64}
P. Hohenberg and W. Kohn, Inhomogeneous electron gas, Phys. Rev. {\bf 136},  B864-871(1964).
\bibitem{Kohn65}
W. Kohn and L. J. Sham, Self-consistent equations including exchange and correlation effects, Phys. Rev. {\bf 140}, A1133-1138 (1965).
\bibitem{Jones89}
R. Jones and O. Gunnarsson, The density functional formalism, its applications and prospects, Rev. Mod. Phys. {\bf 61}, 689-746 (1989).

\bibitem{Hong13}
Q.-J. Hong and A. van de Walle, Solid-liquid coexistence in small systems: A statistical method to calculate melting temperatures, J. Chem. Phys. {\bf 139}, 094114 (2013).
\bibitem{Hong16}
Q.-J. Hong and A. van de Walle, A user guide for \textit{SLUSCHI}: solid and liquid in ultra small coexistence with hovering interfaces, Calphad. {\bf 52}, 88-97 (2016).

\bibitem{Hong15}
Q.-J. Hong, S. V. Ushakov, A. Navrotsky, and A. van de Walle, Combined computational and experimental investigation of the refractory properties of La$_2$Zr$_2$O$_7$, Acta Mater. {\bf 84}, 275-282 (2015).
\bibitem{Hong_HfTaC}
Q.-J. Hong and A. van de Walle, Prediction of the material with highest known melting point from \textit{ab initio} molecular dynamics calculations, Phys. Rev. B {\bf 92}, 020104 (2015).
\bibitem{Mil15}
L. Miljacic, S. Demers, Q.-J. Hong and A. van de Walle, Equation of state of solid, liquid and gaseous tantalum from first principles, CALPHAD: Comput. Coupling Phase Diagrams Thermochem. {\bf 51}, 133-143 (2015).
\bibitem{Guren17}
M. G. Guren, \textit{Ab initio} molecular dynamics simulations of melting phase relations in the system CaO-MgO-SiO$_2$ at pressures of the EarthÕs lower mantle, MS Thesis, University of Oslo, 2017. 
\bibitem{Addington16}
C. K. Addington, Molecular simulation of phase behavior, interfacial phenomena, and pressure effects in porous media, Doctoral dissertation, North Carolina State University, 2016. 

\bibitem{VASP}
G. Kresse and J. Furthm$\ddot{\text{u}}$ller, Efficiency of ab-initio total energy calculations for metals and semiconductors using a plane-wave basis set, Comp. Mater. Sci. {\bf 6}, 15-50 (1996).
\bibitem{BLOCHL94}
P. Bl\"{o}chl, Projector augmented-wave method, Phys. Rev. B {\bf 50}, 17953-17979 (1994).
\bibitem{PBE96}
J. P. Perdew, K. Burke, and M. Ernzerhof, Generalized gradient approximation made simple, Phys. Rev. Lett. \textbf{77}, 3865-3868 (1996).
\bibitem{WIEN2K}
P. Blaha, K. Schwarz, G. Madsen, D. Kvasnicka and J. Luitz, WIEN2k: An Augmented Plane Wave Plus Local Orbitals Program for Calculating Crystal Properties, Institute of Physical and Theoretical Chemistry, TU Vienna, 2001.

\bibitem{SM}
See Supplemental Material at link\_to\_be\_added for details on melting point calculations, comparison of VASP and WIEN2K calculations, and radial distribution functions.

\bibitem{VASP96-2}
G. Kresse, and J. Furthm\"{u}ller, Efficient iterative schemes for ab initio total-energy calculations using a plane-wave basis set. Phys. Rev. B \textbf{54}, 11169-11186 (1996).
\bibitem{VASP99}
G. Kresse, and D. Joubert, From ultrasoft pseudopotentials to the projector augmented-wave method. Phys. Rev. B \textbf{59}, 1758-1775 (1999).

\bibitem{Nose84}
S. Nos\'{e}, A molecular-dynamics method for simulations in the canonical ensemble. Mol. Phys. \textbf{52}, 255-268 (1984).
\bibitem{Nose84-2}
S. Nos\'{e}, A unified formulation of the constant temperature molecular-dynamics methods. J. Chem. Phys. \textbf{81}, 511-519 (1984).
\bibitem{Hoover85}
W. G. Hoover, Canonical dynamics - equilibrium phase-space distributions. Phys. Rev. A \textbf{31}, 1695-1697 (1985).
\bibitem{Klein92}
G. J. Martyna, M. L. Klein, and M. Tuckerman, Nose-Hoover chains - the canonical ensemble via continuous dynamics. J. Chem. Phys. \textbf{97}, 2635-2643 (1992).


\bibitem{Naumov15}
I. I. Naumov and R. J. Hemley, Origin of transitions between metallic and insulating states in simple metals. Phys. Rev. Lett. \textbf{114}, 156403 (2015).
\bibitem{Rousseau11}
B. Rousseau, Y. Xie, Y. Ma, and A. Bergara, Exotic high pressure behavior of light alkali metals, lithium and sodium. Eur. Phys. J. B \textbf{81}, 1 (2011).
\bibitem{Naumov15-2}
 I. I. Naumov, R. J. Hemley, R. Hoffmann, and  N. W. Ashcroft, Chemical bonding in hydrogen and lithium under pressure. J. Chem. Phys. \textbf{143}, 064702 (2015).

\bibitem{Luedemann68}
H. D. Luedemann and G. C. Kennedy, Melting curves of lithium, sodium, potassium and rubidium to 80 kilobars, J. Geophys. Res. \textbf{73}, 2795-2805 (1968).
\bibitem{Mirwald76}
P. W. Mirwald and G. C. Kennedy, Melting temperature of lead and sodium at high pressures, J. Phys. Chem. Solids \textbf{37}, 795-797 (1976).
\bibitem{Boehler85}
C.-S. Zha and R. Boehler, Melting of sodium and potassium in a diamond anvil cell, Phys. Rev. B \textbf{31}, 3199-3201(1985).

\bibitem{Mao05}
E. Gregoryanz, O. Degtyareva, M. Somayazulu, R. J. Hemley, and H. K. Mao, Melting of dense sodium, Phys. Rev. Lett. \textbf{94}, 185502 (2005).

\bibitem{Sugino95}
O. Sugino and R. Car, \textit{Ab initio} molecular dynamics study of first-order phase transitions: melting of silicon, Phys. Rev. Lett. \textbf{74}, 1823-1826 (1995).


\bibitem{Desjarlais13}
M. P. Desjarlais, First-principles calculation of entropy for liquid metals, Phys. Rev. E \textbf{88}, 062145 (2013).
\bibitem{Lin03}
S.-T. Lin, M. Blanco, and W. A. Goddard, III, The two-phase model for calculating thermodynamic properties of liquids from molecular dynamics: Validation for the phase diagram of Lennard-Jones fluids, J. Chem. Phys. \textbf{119}, 11792-11805 (2003).

\bibitem{Alfe11}
D. Alf\`{e}, C. Cazorla, and M. J. Gillan, The kinetics of homogeneous melting beyond the limit of superheating, J. Chem. Phys. \textbf{135}, 024102 (2011).


\bibitem{Urtiew77}
P. A. Urtiew and R. Grover, The melting temperature of magnesium under shock loading, J. Appl. Phys. {\bf 48}, 1122-1126 (1977).

\bibitem{Stinton14}
G. W. Stinton, S. G. MacLeod, H. Cynn, D. Errandonea, W. J. Evans, J. E. Proctor, Y. Meng, and M. I. McMahon, Equation of state and high-pressure/high-temperature phase diagram of magnesium, Phys. Rev. B {\bf 90}, 134105 (2014).

\bibitem{Shaner84}
J. W. Shaner, J. M. Brown, and R. G. McQueen, \textit{High Pressure in Science and Technology}, (North Holland, Amsterdam, 1984).
\bibitem{Boehler97}
R. Boehler and M. Ross, Melting curve of aluminum in a diamond cell to 0.8 Mbar: implications for iron, Earth Planet. Sci. Lett. \textbf{153}, 223 (1997).

\bibitem{Sjostrom16}
T. Sjostrom, S. Crockett, and S. Rudin, Multiphase aluminum equations of state via density functional theory, Phys. Rev. B {\bf 94}, 144101 (2016).
\bibitem{Polsin17}
D. N. Polsin, D. E. Fratanduono, J. R. Rygg, A. Lazicki, R. F. Smith, J. H. Eggert, M. C. Gregor, B. H. Henderson, J. A. Delettrez, R. G. Kraus, P. M. Celliers, F. Coppari, D. C. Swift, C. A. McCoy, C. T. Seagle, J.-P. Davis, S. J. Burns, G. W. Collins, and T. R. Boehly, Measurement of body-centered-cubic aluminum at 475 GPa D., Phys. Rev. Lett. {\bf 119}, 175702 (2017).

\bibitem{Elkins06}
L. T. Elkins-Tanton, \textit{Jupiter and Saturn}, (Chelsea House, New York, 2006). 


\end{thebibliography}
\end{document}